\documentclass[usenatbib,usegraphicx,useAMS,a4]{mn2e}
\usepackage{times}

\newcommand{\kmsec}{\,\mbox{$\mbox{km}\,\mbox{s}^{-1}$}}
\newcommand{\msolyear}{\mbox{$\rm{M}_{\odot}\, \rmn{yr}^{-1}$}}

\author[N.H. Symington, T.J. Harries, R. Kurosawa and T. Naylor]{Neil H. Symington, Tim J. Harries, Ryuichi Kurosawa and Tim Naylor\\
  School of Physics, University of Exeter, Stocker Road, Exeter EX4
  4QL}

\title{T Tauri stellar magnetic fields: He\,{\sc i} measurements}

\begin{document}

\maketitle

\begin{abstract}

We present measurements of the longitudinal magnetic field in the circumstellar
environment of seven classical T~Tauri stars. The measurements are based on
high-resolution circular spectropolarimetry of the He\,{\sc i}~$\lambda 5876$
emission line, which is thought to form in accretion streams controlled by a
stellar magnetosphere.  We detect   magnetic fields in BP~Tau, DF~Tau and
DN~Tau, and detect statistically significant fields in GM~Aur and RW~Aur~A at
one epoch but not at others. We detect no field for DG~Tau and GG~Tau, with the
caveat that these objects were observed at one epoch only. Our measurements for
BP~Tau and DF~Tau are consistent, both in terms of sign and magnitude, with
previous studies, suggesting that the characteristics of T~Tauri magnetospheres
are persistent over several years. We observed the magnetic field of BP~Tau to
decline monotonically over three nights, and have detected a peak field of
4\,kG in this object, the highest magnetic field yet observed in a T~Tauri
star. We combine our observations with results from the literature in order to
perform a statistical analysis of the magnetospheric fields in BP~Tau and
DF~Tau. Assuming a dipolar field, we determine a polar field of $\sim 3$\,kG
and a dipole offset of 40\degr\ for BP Tau, while DF~Tau's field is consistent
with a polar field of $\sim -4.5$\,kG and a dipole offset of 10\degr. We
conclude that many classical T~Tauri stars have circumstellar magnetic
fields that are both strong enough and sufficiently globally-ordered to sustain
large-scale magnetospheric accretion flows.  \end{abstract} 

\begin{keywords}
accretion, accretion discs --
stars: circumstellar matter --
stars: magnetic fields --
stars: pre-main-sequence
\end{keywords}

\section{Introduction}
\label{sec:intro}

It has been postulated that accretion in low-mass pre-main-sequence
stars is magnetically-controlled
\citep[e.g.][]{1991ApJ...370L..39K,1993A&A...274..309C}, with the
magnetic field disrupting a Keplerian disc and the accreting material
plummeting along the field lines onto the stellar surface. This model
provides an attractive solution to the angular momentum dissipation
that is required to produce a slowly rotating protostar, and synthetic
emission line profiles produced assuming a magnetospheric accretion
via a dipolar field are able to reproduce some characteristics of the
observed profiles \citep[e.g.][]{1994ApJ...426..669H,1998ApJ...492..743M}.

One of the key requirements of this model is the presence of a strong,
relatively structured, magnetic field a few stellar radii above the star.
Traditional methods for measuring surface magnetic fields rely on the Zeeman
effect. Actual splitting of absorption line profiles is rarely observed,
because other broadening mechanisms (rotation, pressure, turbulence etc)
usually dominate over the relatively weak Zeeman effect. Nonetheless, if the
intrinsic shape of the profile is well understood, it is possible to attribute
the residual broadening to Zeeman splitting, and hence measure a magnetic field
strength. This method has been used with some success on classical T~Tauri
stars (CTTSs), and surface fields of a few kilogauss have been determined
(\citealt{1992ApJ...390..622B,1999A&A...341..768G,1999ApJ...510L..41J}).

Although surface magnetic fields have been measured for a few
CTTSs, it is not clear how the strength of
these fields changes with radius and how ordered they are
(e.g. \citealt{1998ApJ...494..336S}). Measurements of circumstellar
magnetic fields require observations of the Zeeman effect in an
emission line, and here the Zeeman broadening technique fails --
primarily because the lines are broadened by the bulk motion of the
accreting gas, and the intrinsic profile of the line cannot be
determined with any certainty. Fortunately Zeeman splitting also
results in a polarization signature, with the $\sigma$-components of
the transition being circularly polarized in opposite senses. The
separation between these components is directly proportional to the
mean longitudinal field strength \citep{1947ApJ...105..105B}.

\citeauthor{1986MNRAS.219..927J} (\citeyear{1986MNRAS.219..927J},
hereafter JP) attempted to measure the magnetic fields in three
T~Tauri stars using this effect, and found a 2.3-$\sigma$ detection ($-548$G)
for RU~Lup, which was not confirmed by later observations
\citep{1987MNRAS.227..797J}. More recently
\citet{1999ApJ...510L..41J}  measured the Zeeman shift in
circular polarization of the He\,{\sc i}~$\lambda 5876$ in BP~Tau,
using an echelle spectrograph on a 2.7-m telescope at McDonald
Observatory. The He\,{\sc i}~$\lambda 5876$ line is thought to
form above hot impact regions on the stellar surface in streams of gas being accreted by the star (e.g. \citealt{1993AJ....106.2024V}).
\citet{1994AJ....108.1056E} found that the line showed an inverse P~Cygni
profile for stars with high levels of continuum veiling -- two well-known 
accretion indicators. 
\citet{1999ApJ...510L..41J} inferred a mean longitudinal magnetic field of 2.4\,kG, which was the first
direct evidence of magnetic accretion in a CTTS. Similar observations
on three more CTTSs were presented by \cite{2001csss...11..521J} and
\cite{2003csss...12..729V}. 

Here we present measurements of the longitudinal magnetic field in
seven CTTSs using the He\,{\sc i}~$\lambda 5876$ emission line. The
aim of these observations was to expand the sample of CTTSs with known
fields and to examine the long-term (many rotation cycles) stability
of the magnetosphere.

\section{Observations and data reduction}
\label{sec:obsanddr}

\subsection{Zeeman effect}
\label{sec:zeeman}

In this paper we exploit the Zeeman effect to measure the magnetic
field around CTTSs. If there is a net magnetic
field along the observer's line of sight, a spectral line is split and the
separation in wavelength between the two $\sigma$-components is a measure of
the field strength. The wavelength shift is
\begin{equation}
  \label{eqn:dlambda_sym}
\Delta\lambda = 2\, \frac{e}{4 \pi m_e c^2}\, \lambda^2\, g_{\mathrm{eff}}\, B_z
\end{equation}

\begin{equation}
  \label{eqn:dlambda_num}
  \phantom{\Delta\lambda } = 9.34\times 10^{-7}\, \lambda^2 g_{\mathrm{eff}}\, B_z\, \textrm{m\AA}\, \textrm{kG}^{-1}
\end{equation}
  where $\lambda$ is the rest wavelength of line (in \AA),
  $g_{\mathrm{eff}}$ is the effective Land\'{e} $g$-factor of the line
  transition and $B_{z}$ is the net longitudinal magnetic field
  \citep{1991A&AS...89..121M}.
  The value of $g_{\mathrm{eff}}$ adopted for He\,{\sc i}~$\lambda 5876$ was $1.11$, but the line is formed by multiple transitions and $g_{\mathrm{eff}}$ is dependent on the environment (see e.g. \citealt{1999ApJ...510L..41J}). 
  For other spectral lines, Land\'{e} $g$-factors were obtained from the solar
line list table of J.F.~Donati reproduced at
\texttt{http://bass2000.bagn.obs-mip.fr}.

\subsection{Observations}
\label{sec:obs}

Spectra were obtained with the ISIS spectrograph at the Cassegrain
focus of the 4.2-m William Herschel Telescope. The configuration for
circular spectropolarimetry is described in \citet{IsisPolManual}. A
quarterwave plate first converts from circular to linear
polarization. Light from the target object then passes through one
aperture of a comb dekker; the other apertures sample the background
sky emission.  A comb dekker with 20-arcsec apertures was used for all
exposures except the tungsten lamp flat-fields. The star and sky beams
all pass through a Savart plate which spatially separates the
orthogonal linear polarizations into o- and e-beams, each of which
then passes through the standard spectrograph optics and is recorded
on a CCD detector. Rotating the quarterwave plate between exposures,
by $90\degr$ relative to the first position, swaps the polarization
sense of each beam and should reverse direction of the measured
wavelength shift. Many of the possible systematic errors in the optics
or detector would lead to unequal inferred magnetic field magnitudes
from a pair of exposures, and this is the first of several tests we
can use to assess the veracity of our results.
 
The observations were made on nights beginning 2001 December 3--5 and 2002
December 21--23 with the R1200 grating on the ISIS red
arm. Approximately half of each run was lost to weather or technical
problems. We observed calibration targets, and seven bright CTTSs in the 
Taurus-Auriga region with known He\,{\sc i} emission lines \citep{1994AJ....108.1056E}.    
The TEK4 CCD detector was used during the 2001 run, providing
$0.40\rm{\AA}$ per pixel reciprocal dispersion; the newer MARCONI2 CCD
was employed in 2002 to obtain $0.22\rm{\AA}$ per pixel. The
resolution achieved was 0.7\AA\ (FWHM of the arc lines). Maximum
sensitivity of the system was selected by inserting a
circular-polarizing filter during setup, and finding the
quarterwave-plate angle that gave the maximum contrast between the
intensity of the o- and e-beams. Observations of the target objects
(typically 1000-s per frame) were bracketed by exposures with
copper-neon and copper-argon comparison lamps to provide a wavelength
calibration.

\subsection{Data reduction}
\label{sec:reduction}
The procedure for obtaining a magnetic field measurement requires
cross-correlation of two spectra recorded simultaneously on different
areas of the CCD. Differential errors in the extraction or calibration
of the spectra will create a false magnetic field detection; it is
important to test the systematic effects are well below the (less than
one pixel) expected Zeeman wavelength shift.

Each CCD frame was first bias-subtracted and flat-fielded. The
flat-field frames had a large intensity gradient towards the edges
where vignetting was significant, so these frames were normalised so
that they had a value close to 1 across the central area, where the
spectral lines of interest were recorded. In the 2002 run, flat-field
frames were recorded at both quarterwave-plate rotation positions
($\theta_0$,\ $\theta_0 + 90\degr$) and the appropriate master
flat-field used for correcting each of the target exposures in a
pair. Similarly, the arc comparison frames were recorded at both
quarterwave-plate angles, and the appropriate arc spectrum was used for
calibrating each target spectra. 

The 2001 observations contained calibration frames made at only one rotation
angle of the quarter-wave plate, and it was necessary to test the effect of
reducing data from both quarter-wave plate positions with a flat-field from
just one position. Test reductions were carried out with the 2002 data using mismatching
flat-fields and the derived wavelength shifts were not significantly different from the results using flat-fields from
both quarter-wave plate positions.
These results were also unaffected when a single arc-lamp exposure was used to
calibrate each complementary pair of target exposures, so we had confidence in
using these techniques for the 2001 data.  

For each exposure, the two stellar spectra (corresponding to the left- and
right-handed polarisation states) were extracted, as well as two spectra
sampling the sky background. The sky spectra were of very much lower intensity
than the stellar continuum and were unstructured over the wavelength extent of
the He\,{\sc i} $\lambda5876$ line. Tests on sample frames showed that sky
subtraction did not affect the results; it was decided that this process only
contributed noise and should be omitted from the data reduction sequence.    

Spectra of the arc lamps were obtained from their frames using the same
extraction windows used for the stellar spectra, then a third-order polynomial
wavelength scale was fitted using identifiable emission lines as references.
The typical RMS error from the fit to the arc lines was $\sim0.03$\AA.
 The four stellar spectra (an o- and an
e-beam recorded at each quarterwave plate position) were rebinned to a common
linear wavelength scale in preparation for determining the wavelength shift. The scale
was chosen to be similar to that of the individual spectra. 


\subsection{Cross-correlation analysis}
\label{sec:xcorr}
A range of wavelength bins containing the emission or absorption line
of interest were copied from the full spectra for cross-correlation
using the procedure described in JP. We manually inspected the spectra
to choose the limits for the line region, rather than employ an
automated procedure. The emission lines in the T Tauri-class spectra
were generally strong compared to the continuum, so most of the
photons were recorded in a narrow wavelength range and it was not felt
necessary to include the line wings where it was difficult to identify
the transition to the local continuum intensity. 

The two rebinned spectra from each CCD frame were cross-correlated by
evaluating a $\chi^2$ statistic, S, for one spectrum being used as a
model for the other (after allowing for intensity rescaling). As
described in JP, we made whole-pixel shifts between the spectra and
fitted a parabola to the three values of S that were closest to
the minimum value, S(min).  The minimum turning point of the parabola gives the
wavelength shift between the two spectra, $\Delta\lambda$, that is used to obtain the
magnetic field strength $B_{z}$ (Equation~\ref{eqn:dlambda_num}). Table~\ref{table:obs} lists $B_{z}$ for each of the target exposures. 

Although we had propagated uncertainty
estimates for each pixel or wavelength bin throughout the data
reduction, when the spectra were rebinned to a common wavelength scale
it became non-trivial to estimate the uncertainty contribution from
each source bin to each target bin because the errors are then
correlated. We therefore used the ratio of variances method (described
in JP; \citealt{1976ApJ...208..177L}) to estimate the 1-$\sigma$
uncertainty interval for each Zeeman wavelength shift.  The interval
is found from the cross-correlation shift $\chi^2$ statistic, S,
relative to S(min), its value at the minimum turning point of the parabola, such that:

\begin{equation}
  \label{eqn:Sstat}
 \mathrm{S} < \mathrm{S(min)} \left[ 1 + \frac{\mathrm{p}}{\mathrm{N}'-\mathrm{p}} F(\mathrm{p},\, \mathrm{N}'-\mathrm{p}, \,\alpha)   \right]       
\end{equation}

where $\mathrm{p}$ is the number of free parameters used to find S(min);
$\mathrm{N}'$ is the number of independent wavelength bins in the
spectra; $\alpha$ is the confidence level required (i.e. $\sim0.68$
for the 1-$\sigma$ interval); and $ F(\mathrm{p},\,
\mathrm{N}'-\mathrm{p}, \,\alpha) $ is the Fischer-Snedecor ratio of
variances function.  One spectrum from each pair was scaled in
intensity to match the other, and then shifted in wavelength, so
$\mathrm{p} = 2$. We follow the advice of JP, who show that after
rebinning the spectra, the effective number of independent wavelength
bins is reduced to $5/6$ of the number of bins in the raw spectra. The interval in Equation~\ref{eqn:Sstat} defines a range of wavelength shifts that gives (with Equation~\ref{eqn:dlambda_num}) the 1-$\sigma$ uncertainties on the $B_z$ values presented in Table~\ref{table:obs}.

A hybrid scheme was also tried which used
inverse variance weighting when calculating the $\chi^2$
statistics. The advantage of this alternative was that the variance
associated with each wavelength bin included information about the
flat-field uncertainties, detector noise etc. The results
were not significantly altered by this change in procedure, and we
concluded that the original cross-correlation method was adequate.

We found that small changes made to our procedures (such as altering the
wavelength limits of the emission line, or the trials using single
calibration frames in Section~\ref{sec:reduction}) typically perturbed the
individual measurements by approximately $100$G. When the uncertainties in the
polarization standards (Section~\ref{sec:standards}) are also taken into
account, we estimate that the systematic errors in our results are probably of
magnitude $\sim200$G and we would not confidently claim to detect magnetic
fields of lesser strength. We have not attempted to account for the systematic
effects when quoting uncertainty intervals in this paper; they are always the
formal errors derived from the cross-correlation procedure.  

\begin{figure*}
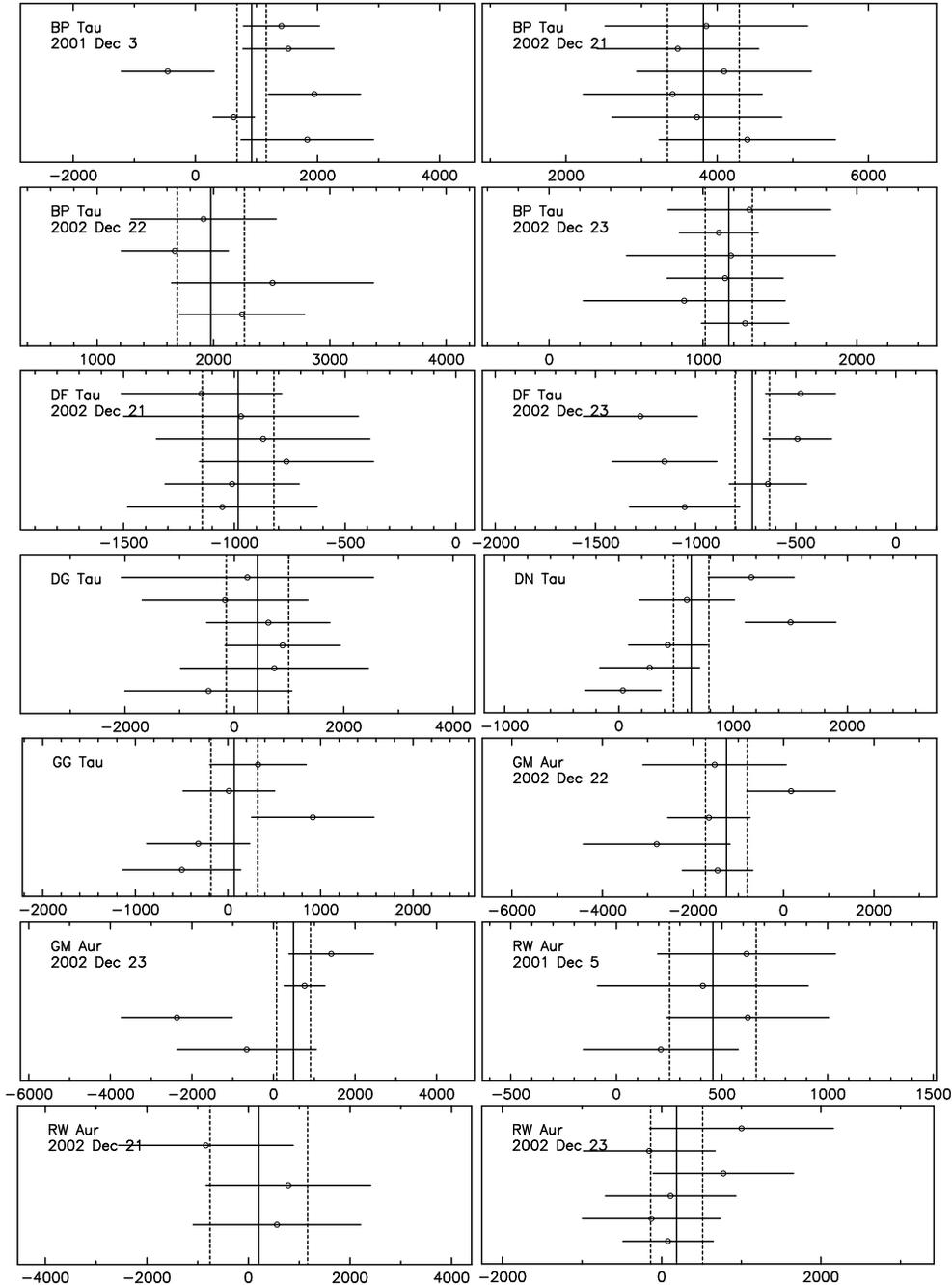

\caption{Plots of the $B_z$ (G) results from individual exposures. The circles are the measurements derived from individual exposures; the horizontal lines are the 1-$\sigma$ uncertainty intervals for those values. Each night's data for each object were combined using an inverse-variance weighted mean. The vertical solid and dashed lines show the mean and 1-$\sigma$ intervals respectively.}
\includegraphics[angle=270,width=6.3cm]{fig1_1a.eps}
\includegraphics[angle=270,width=6.3cm]{fig1_1b.eps}
\includegraphics[angle=270,width=6.3cm]{fig1_2a.eps}
\includegraphics[angle=270,width=6.3cm]{fig1_2b.eps}
\includegraphics[angle=270,width=6.3cm]{fig1_3a.eps}
\includegraphics[angle=270,width=6.3cm]{fig1_3b.eps}
\includegraphics[angle=270,width=6.3cm]{fig1_4a.eps}
\includegraphics[angle=270,width=6.3cm]{fig1_4b.eps}
\includegraphics[angle=270,width=6.3cm]{fig1_5a.eps}
\includegraphics[angle=270,width=6.3cm]{fig1_5b.eps}
\includegraphics[angle=270,width=6.3cm]{fig1_6a.eps}
\includegraphics[angle=270,width=6.3cm]{fig1_6b.eps}
\includegraphics[angle=270,width=6.3cm]{fig1_7a.eps}
\includegraphics[angle=270,width=6.3cm]{fig1_7b.eps}
\label{fig:bzplots}
\end{figure*}

The magnetic field of a target T~Tauri star was expected to vary little over
the duration of the observations (less than $\sim 1$h) on a given night. We
therefore obtained a nightly $B_z$ measurement for each star by computing a
mean value weighted by the inverse variance of the individual estimates.
Fig.\ref{fig:bzplots} shows both the individual measurements and the mean
values, with their 1-$\sigma$ uncertainty intervals.   We present the combined
results in Table~\ref{table:results}, along with characteristic stellar data
from the literature. 

To test our estimates of the 1-sigma uncertanties we can calculate a
reduced chi-squared statistic for all our measurements, with respect to the mean for
the object on the night in question.  This yields a $\chi_{\nu}^2$ of 0.836 with
59 degrees of freedom.  Since we expect 81 per cent of such experiments to
exceed this value, it suggests that both our assumption that the
polarization is unchanging within the $\sim$1h of our observations, and our
calculation of the uncertainties is broadly correct. Table~\ref{table:results} also gives
the values of $\chi_{\nu}^2$ for each object on each night.  Although there are
a couple of uncomfortably high values of $\chi_{\nu}^2$ for individual objects
on individual nights, given the above result, and the low number of
degrees of freedom for the nightly chi-squareds, their statistical
significance is unclear.  Furthermore, the highest value corresponds to
the data with the smallest uncertainty (86G) and thus we would expect it
to be affected by the systematic uncertainties that we estimate to be
about 200G.  We therefore conclude that the $\chi^2$ analysis supports our
conclusion that there are systematic effects at the 200G level, but above
this level our uncertainties are probably correct.

\subsection{Test stars}
\label{sec:standards}

We observed three targets with dependable levels of circular
polarization to test our instrumental setup and data reduction
process. The Sun has a magnetic field strength much smaller than the
sensitivity of our equipment, so it was used as a null
reference. Sunlight reflected from Vesta provided practical access to
the solar spectrum; for the purposes of this experiment, the asteroid
can be considered to be a passive reflector of circularly-polarized
light. In poor seeing conditions, only two measurements had reasonable
signal-to-noise ratios and these two useful exposures were obtained
with the quarterwave plate at the same angle, so we can not test that
the same magnitude of wavelength shift is measured as the beams are
reversed (as described in Section~\ref{sec:reduction}). An
inverse-variance weighted mean of the shift in the Na~\textsc{i}
5889\AA\ line represented a mean field strength $\mathrm{B}_z = -20
\pm 180$G, which is consistent with the expected null result.

\begin{figure}
\includegraphics[width=8cm,clip=true]{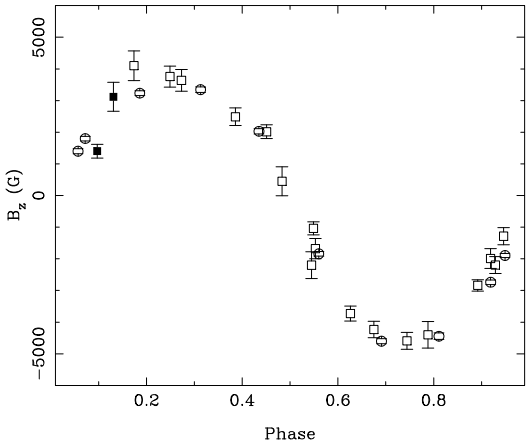}
\caption{Magnetic field measurements of the Ap star 53~Cam. Our data are shown as filled squares,
  along with those of  \citet{1998MNRAS.297..236H} (open squares) and
   \citet{2000MNRAS.313..851W} (open circles).}
\label{fig:53cam}
\end{figure}

Observations of the magnetic Ap star 53 Cam permitted comparison with
measurements published by other authors. Data obtained in both the
2001 and 2002 runs were phased using the ephemeris of
\citet{1998MNRAS.297..236H}.
Our two measurements are shown in Fig.~\ref{fig:53cam}, along with
those from two other sources.  Rather than choose a sign convention
for $B_z$ based on the spectrograph setup, we have assumed that our 53
Cam measurements are consistent with the previously published data,
and should therefore both have a positive sign. Our data then are of the magnitude expected from the previous studies, indicating that our
measurement technique is reliable. The sign convention adopted is
later shown to also be consistent with previous measurements of
T~Tauri magnetic fields published by other authors.

Observations were made of a linear polarization standard star to determine
whether our instrumentation would apparently detect a Zeeman wavelength shift
because of unwanted conversion from linear and circular polarization in the
optics. The star HD25443 (5.1 per cent \textit{V}
band linear polarization; \citealt{1990AJ.....99.1243T}) was observed during the 2001 run and
several spectral lines were cross-correlated. The wavelength shift found by our
data reduction process would be equivalent to a magnetic field of $-194 \pm
45$G, which would be a significant detection. It is possible that this star
does have an intrinsic longitudinal magnetic field, but it is more likely that
the apparent wavelength shift between the two beams is caused by imperfect
polarization optics.  We expect that any effects of polarization cross-talk are
going to be within the 200G systematic error estimate we discussed above and do
not threaten the validity of our experimental results.

\begin{table*}

  \caption{\label{table:obs} Summary of T~Tauri star observations and individual results.
  Times quoted are UT at exposure midpoint. Quarterwave plate rotation angles
  ($\phi$) are quoted in degrees, relative to the adopted zero point
  (Section \ref{sec:obs}). Uncertainties in the longitudinal magnetic field,
  $B_z$, are the 1-$\sigma$ interval. }
  
\tiny{
  \begin{tabular}{cccrrrrccccrrrr}
Name & Night & JD$-$2452000 & $\phi$ & $\Delta\lambda$ (m\AA) & $B_z$(G) & $\pm$ 1-$\sigma$ & \phantom{}&
Name & Night & JD$-$2452000 & $\phi$ & $\Delta\lambda$ (m\AA) & $B_z$(G) & $\pm$ 1-$\sigma$ \\
\hline
BP Tau & 2001 Dec 03 & 247.476 & 0 & 64 & 1800 & 1100 & &  DN Tau & 2002 Dec 23 & 632.317 & 90 & 53 & 1490 & 410 \\ 
       &             & 247.488 & 90 & 23 & 630 & 340      & &      &             & 632.329 & 0 &$-0.4$& $-10$ & 440 \\
       &             & 247.500 & 0  & 70 & 1950 & 750      & &      &             & 632.341 & 90 & 21 & 600 & 420 \\
       &             & 247.512 & 90 &$-16$& $-450$ & 760    & &      &             & 632.353 & 0 & 10 & 270 & 440 \\
       &             & 247.525 & 0  & 54 & 1520 & 750      & &      &             & 632.369 & 90 & 42 & 1160 & 380 \\
       &             & 247.537 & 90 & 50 & 1410 & 622     & &      &             & 632.381 & 0 & 15 & 430 & 350 \\  
       & 2002 Dec 21 & 630.385 & 0  & 160 & 4400 & 1200    & &  GG Tau & 2002 Dec 23 & 632.401 & 0 &$-18$&$ -500$ & 640 \\
       &             & 630.391 & 90 & 130 & 3700 & 1100   & &      &             & 632.412 & 90 & 33 & 920 & 660 \\  
       &             & 630.412 & 0  & 120 & 3400 & 1200    & &      &             & 632.424 & 90 & 0.4 & 10 & 500 \\   
       &             & 630.424 & 90 & 150 & 4100 & 1200   & &      &             & 632.436 & 0 &$-11$& $ -320$ & 560 \\
       &             & 630.441 & 0  & 130 & 3500 & 1100    & &      &             & 632.450 & 90 & 12 & 330 & 520 \\  
       &             & 630.453 & 90 & 140 & 3900 & 1300   & &  GM Aur & 2002 Dec 22 & 631.688 & 90 &$-100$&$ -2800$ & 1600 \\       
    & 2002 Dec 22 & 631.633 & 90 & 90 & 2510 & 870    & &      &             & 631.700 & 0 &$-52$& $ -1460$& 780 \\   
    &             & 631.645 & 0 & 81 & 2250 & 540     & &      &             & 631.712 & 0 &$-59$&$ -1650$ & 910 \\  
    &             & 631.657 & 0 & 60 & 1670 & 460     & &      &             & 631.724 & 90 & 6 & 170 & 980 \\     
    &             & 631.668 & 90 & 68 & 1910 & 620    & &      &             & 631.737 & 90 &$-54$&$ -1500$ & 1600 \\
    & 2002 Dec 23 & 632.564 & 0  & 45 & 1270 & 290     & &      & 2002 Dec 23 & 632.726 & 0 & $-25$ &$ -700$ & 1700 \\  
    &             & 632.575 & 90 & 32 & 880 & 660     & &      &             & 632.738 & 90 & 27 & 760 & 501 \\     
    &             & 632.587 & 0  & 41 & 1140 & 380     & &      &             & 632.747 & 90 & 50 & 1400 & 1100 \\   
    &             & 632.599 & 90 & 42 & 1180 & 680    & &      &             & 632.753 & 0 &$-86$& $ -2400$ & 1400 \\ 
    &             & 632.612 & 0  & 39 & 1100 & 260     & &  RW Aur A & 2001 Dec 05 & 249.700 & 0 & 8 & 210 & 370 \\ 
    &             & 632.624 & 90 & 47 & 1300 & 530    & &      &            & 249.711 & 90 & 22 & 620 & 380 \\      
DF Tau & 2002 Dec 21 & 630.474 & 0 &$-38$& $-1050$ & 430 & &      &            & 249.723 & 0 & 15 & 410 & 500 \\     
    &             & 630.491 & 90 &$-36$& $-1010$ & 300   & &      &            & 249.735 & 90 & 22 & 620 & 420 \\    
    &             & 630.506 & 0  &$-28$& $ -770$ & 400     & &      & 2002 Dec 21 & 630.659 & 0 & 21 & 600 & 1700 \\   
    &             & 630.520 & 90 &$-31$& $ -870$ & 480    & &      &             & 630.671 & 90 & 29 & 800 & 1600 \\  
    &             & 630.532 & 0  &$-35$& $ -970$ & 530     & &      &             & 630.683& 0 &$-29$&$ -800$ & 1700 \\
    &             & 630.544 & 90 &$-41$ & $ -1150$ & 360   & &      & 2002 Dec 23 & 632.484 & 90 & 28 & 780 & 880 \\   
    & 2002 Dec 23 & 632.649 & 90 &$-23$& $ -640$  & 190   & &      &             & 632.497 & 0 & 3 & 80 & 570 \\     
    &             & 632.661 & 0  &$-38$& $-1060$ & 280    & &      &             & 632.509 & 90 &$-6$&$ -160$ & 830 \\
    &             & 632.672 & 0  &$-42$& $-1160$ & 260    & &      &             & 632.521 & 0 &$-5$&$ -130$ & 870 \\ 
    &             & 632.684 & 90 &$-18$& $-490$ & 170    & &      &             & 632.533 & 90 & 36 & 1000 & 1200 \\ 
    &             & 632.696 & 90 &$-17$& $ -480$ & 170    & &      &             & 632.545 & 0 & 4 & 110 & 820 \\    
    &                & 632.708 & 0 &$-46$& $ -1280$ & 290    \\
DG Tau & 2002 Dec 21 & 630.564 & 0 &$-18$& $ -500$ & 1500 \\
    &                & 630.576 & 90 & 25&700 & 1700    \\
    &                & 630.588 & 0  & 32 &880 & 1100     \\
    &                & 630.600 & 90 & 22 & 620 & 1100    \\
    &                & 630.615 & 0  &$-6$& $-170$ & 1500    \\
    &                & 630.627 & 90 & 9 &240 & 2300    \\
\hline
  \end{tabular}
}
\end{table*}

\section{Results}
\label{sec:results}

Here we describe our measurements of individual objects, and review
the observational evidence for magnetically controlled phenomena.

\subsection{Null detections}
\label{sec:null}

Two of the target CTTSs had results that were consistent with zero longitudinal field. Our cross-correlation analysis (Section~\ref{sec:xcorr}) leads to formal 2-$\sigma$ upper limits of $|B_z| < 0.80$\,kG for DG~Tau and $|B_z| < 0.32$\,kG for GG~Tau. Each of these stars was observed on only one night, so it is possible that the measurements were made at epochs when the orientations of the magnetic fields were unfavourable. 

\subsection{Possible detections}
\label{sec:marginal}
\begin{table*}

  \caption{\label{table:results}
  Nightly $B_z$ results for each star. The $v\, \sin i$ data are from \citet{1989AJ.....97..873H}. Other properties for each object were from references: 
  1. \citet{1979ApJ...227L.105C};
  2. \citet{1998ApJ...492..323G};
  3. \citet{1995ApJ...452..736H};
  4. \citet{1993AJ....106.2024V}. Mass accretion rate ($\dot{{\rm M}}$) is in units of $10^{-7}\msolyear$, $\mathrm{v}\, \sin i$ in \kmsec.  }
  
  \begin{tabular}{lllcccrrr}
    Name    &Sp. Type (ref.) & $\mathrm{v}\, \sin i$ & $\dot{{\rm M}}$  (ref.)  & Night beg.   & \# Obs. & $B_z$(G) & $\pm$ 1-$\sigma$ & $\chi^2_{\nu}$\\
\hline
BP Tau  & K7 (1) & 10.0 & 0.288 (2) & 2001 Dec 03 & 6   & 920      &  240             & 1.57 \\ 
        &         &     &           & 2002 Dec 21 & 6   & 3820     &  480             & 0.106 \\ 
        &         &      &          & 2002 Dec 22 & 4   & 1980     &  290             & 0.361 \\ 
        &         &      &          & 2002 Dec 23 & 6   & 1170     &  150             & 0.093 \\ 
DF Tau  & M0.5 (1)& 16   & 1.767 (2)& 2002 Dec 21 & 6   & $-980$     &  160             & 0.121 \\
        &         &      &          & 2002 Dec 23 & 6   &$ -717 $    &  86              & 2.40 \\ 
DG Tau  &K7--M0 (3)& 22   & 20 (3)   & 2002 Dec 21 & 6   & 420      &  570             & 0.151 \\ 
DN Tau  &  M0     & 10.2 & 0.033(2) & 2002 Dec 23 & 6   & 630      &  160             & 2.20 \\ 
GG Tau  &K7--M0 (1)& 10.2 & 0.175(2) & 2002 Dec 23 & 5   & 70       &  250             & 0.795 \\ 
GM Aur  &K7--M0 (1)& 12.4 & 0.096(2) & 2002 Dec 22 & 5   & $-1260 $   &  460             & 0.825 \\ 
        &         &      &          & 2002 Dec 23 & 4   & 490      &  420             & 1.98 \\ 
RW Aur A& K3 (4)  & 19.8 & 3.4 (4)  & 2001 Dec 05 & 4   & 460      &  210             & 0.263 \\ 
        &         &      &          & 2002 Dec 21 & 3   & 210      &  960             & 0.271 \\ 
        &         &      &          & 2002 Dec 23 & 6   & 190      &  330             & 0.258  \\ 
\hline
  \end{tabular}
\end{table*}

In 2001 we detected a magnetic field on RW~Aur ($+0.46\pm0.21$\,kG)
although it was at the limit of our setup's sensitivity. Subsequent
observations on two nights in 2002 found $B_z$ close to zero.
On the night of 2002 December 22, $B_z$ for GM~Aur was found to be $-1.26 \pm
0.46$\,kG -- a significant detection -- but by the following night was
consistent with zero ($+0.49 \pm 0.42$\,kG). 

These stars might be expected to show evidence of magnetospheric accretion.
\citet{1998ApJ...492..323G} found the accretion rate of GM~Aur to be of the
order of $\sim 10^{-8}$\msolyear and \citet{2001ApJ...550..944M} presented fits to the Balmer
lines and Na\,D profiles of RW~Aur based in a magnetospheric accretion model with $\dot{{\rm M}}= \sim 10^{-8}$ \msolyear. High resolution time-series spectroscopy of
RW~Aur~A revealed periodic modulation of the emission line profiles
($P=2.77$\,d, \citealt{2001A&A...369..993P}). Non-axisymmetric accretion, due
to either a close companion or an offset dipole configuration, was postulated
to explain the variability. Magnetic fields are also implicated in the
formation of its jet \citep{1994ApJ...427L..99H}. 

We are less confident in our claims for these CTTSs than for others
(Section~\ref{sec:detections}).  The behaviour may be evidence for rotational
modulation of the magnetic fields, but detections at more than one epoch would
be desirable for these two targets.

\subsection{CTTS with detected magnetic fields}
\label{sec:detections}
\citet{1999ApJ...510L..41J} used circular spectropolarimetry to search
for magnetic fields on BP~Tau. They determined an upper-limit of
200\,G for the surface field, but found $B_z = 2460 \pm 120$\,G in the
magnetosphere from the He\,{\sc i}~$\lambda 5876$ line. They concluded
that the accretion must be confined to streams with small footprints
on the stellar surface. Further He\,{\sc i}~$\lambda 5876$ observations
\citep{2003csss...12..729V} showed that $B_z$ is variable on a
time-scale of days, indicating rotational modulation of a structured
magnetosphere.

Our BP~Tau observations from 2001 and 2002 confirm those of
\citet{1999ApJ...510L..41J} and \citet{2003csss...12..729V}, although
our peak $B_z$, at 4\,kG, is the strongest yet found for any CTTS. In
our sign convention (see Section~\ref{sec:standards}), $B_z$ was
always positive, as it was in the previously published results. The
field strength appeared to be monotonically declining over the three
consecutive nights of the 2002 observing run. Three data points are
not sufficient to draw any conclusions about periodicities in the
data, but the \citet{2003csss...12..729V} observations also showed a
smooth change over six consecutive nights (with a minimum value of
$0.3$\,kG), so it likely that the dominant magnetic time-scale is many
days, possibly matching the stellar rotation period (7.6\,d; \citealt{1990AJ....100.1957S}). We investigate
this possibility further in Section~\ref{sec:stats}.


DF~Tau is a binary system \citep{1990ApJ...357..224C}, although the
primary is responsible for the continuum and line emission in the blue
\citep{2001A&A...372..922L}. \citet{2003csss...12..729V} observed a 
longitudinal field strength
that varied smoothly between $-0.3$ and $-1.0$\,kG over six nights. Our
two measurements are within this range, so we have another indication
that the T~Tauri magnetospheres have persistent long-term
characteristics.

It should be made clear that Fig.~\ref{fig:bzplots} appears to show the
individual DF~Tau measurements from 2002 December 23 alternating between two
separate magnetic field strengths. The bimodal distribution is
accounted for by the change of quarterwave-plate angle between
exposures. The sign of the magnetic field, but not its magnitude are
expected to change after that procedure.  Small-scale magnitude
variations were seen in other targets, but the later set of DF~Tau
observations shows the most significant effect. It is likely that
imperfections in the spectrograph optics are affecting either the
polarization signature of the star, or introducing a systematic error
into our wavelength calibrations. The mean magnitude of $B_z$ in these
data is sufficiently high that we are confident in ruling out a null
detection, but the formal error estimates in the final result should
be treated sceptically.


Our measurements of $B_z$ for DN~Tau show an overall statistically significant
detection ($+0.63 \pm 0.16$\,kG) although several of the contributing
individual data points were consistent with zero magnetic field. We
suggest that we have probably made the first detection of a
longitudinal magnetic field from this star.

Two of the CTTSs with detected magnetic fields have shown evidence for surface
hotspots: DN~Tau (\citealt{1986ApJ...306..199V}; \citealt{1986A&A...158..149B})
and DF~Tau \citep{1993A&A...272..176B} -- which has been mapped by
\citet{1998MNRAS.295..781U} using Doppler tomography. 

\begin{table}

\caption{\label{table:valenti} Published longitudinal magnetic field measurements for the T~Tauri stars BP~Tau and DF~Tau
 using the He\,{\sc i}~$\lambda 5876$ line. References: 1. \citet{2003csss...12..729V};
2. \citet{1999ApJ...510L..41J}. }

\begin{tabular}{llrrc}
Name    & Night       & $B_z$(G) & $\pm$ 1-$\sigma$ & Ref.\\
\hline
BP~Tau  & 1997 Nov 21 & $2460 $  & $120$ & 2 \\
        & 1998 Nov 26 & $ 2740$  & $170$ & 1 \\
        & 1998 Nov 27 & $ 1490$  & $170$ & 1 \\
        & 1998 Nov 28 & $ 850 $  & $150$ & 1 \\
        & 1998 Nov 29 & $ 330 $  & $280$ & 1 \\
        & 1998 Nov 30 & $ 390 $  & $320$ & 1 \\
        & 1998 Nov 31 & $ 610 $  & $150$ & 1 \\
DF~Tau  & 1998 Nov 26 & $ -300$  & $110$ & 1 \\
        & 1998 Nov 27 & $ -520$  & $90 $ & 1 \\
        & 1998 Nov 28 & $ -810$  & $80 $ & 1 \\
        & 1998 Nov 29 & $ -940$  & $90 $ & 1 \\
        & 1998 Nov 30 & $-1020$  & $330$ & 1 \\
        & 1998 Nov 31 & $-1000$  & $110$ & 1 \\
\hline
  \end{tabular}
\end{table}

\subsection{Models for BP~Tau and DF~Tau}
\label{sec:stats}
Our observations of the longitudinal magnetic field of T~Tauri stars,
have some targets in common with those recently published by
\citet{1999ApJ...510L..41J} and \citet{2003csss...12..729V}. Those authors' results for the stars BP~Tau and DF~Tau are shown in Table\ref{table:valenti}.  With all
three data sets, there is a sufficient number of measurements (11 and
8 respectively of BP~Tau and DF~Tau) to permit a preliminary
statistical analysis of the magnetic field properties.

We investigated the case where the measured field arises solely from
an accretion stream following the field lines of a dipolar
magnetosphere. The line-of-sight component of such a magnetic
configuration, $B_z$, is given by:

\begin{equation}
B_z = B_{\mathrm{pole}} ( \cos i\, \cos \beta + \sin i\,\sin\beta\, \cos\phi)
\end{equation}

where $B_{\mathrm{pole}}$ is the field strength at the poles of the
magnetosphere; $i$ is the observer's inclination with respect to the
rotation axis; $\beta$ is the dipole offset -- the inclination of
the magnetic polar axis with respect to the rotation axis; and $\phi$
is the rotational phase angle at the time of observation.

A Monte Carlo computer code was used to make simulated observations at
many phases based on input parameters of $B_{\mathrm{pole}}$, $i$ and
$\beta$. The relative spacing (but not actual values) of the phase
points were chosen to reflect the timing of the real observations,
i.e. a random phase point was first chosen to represent `night 1' of
an observing run, and then the elapsed time to the next observation,
divided by the rotation period, was used to derive the subsequent
phase points. The periods used were 7.6\,d \citep{1990AJ....100.1957S},
for BP~Tau and 8.5\,d for DF~Tau \citep{1990AJ.....99..946B}. The
inclination parameter, $i$, was allowed to vary by 10\degr about fixed
values from the literature: 50$\degr$ for BP~Tau, 70$\degr$ for DF~Tau
(\citealt{2003csss...12..729V}).
\begin{figure*}
\centering
\begin{tabular}{c}
\includegraphics[width=180mm,angle=0,clip=true]{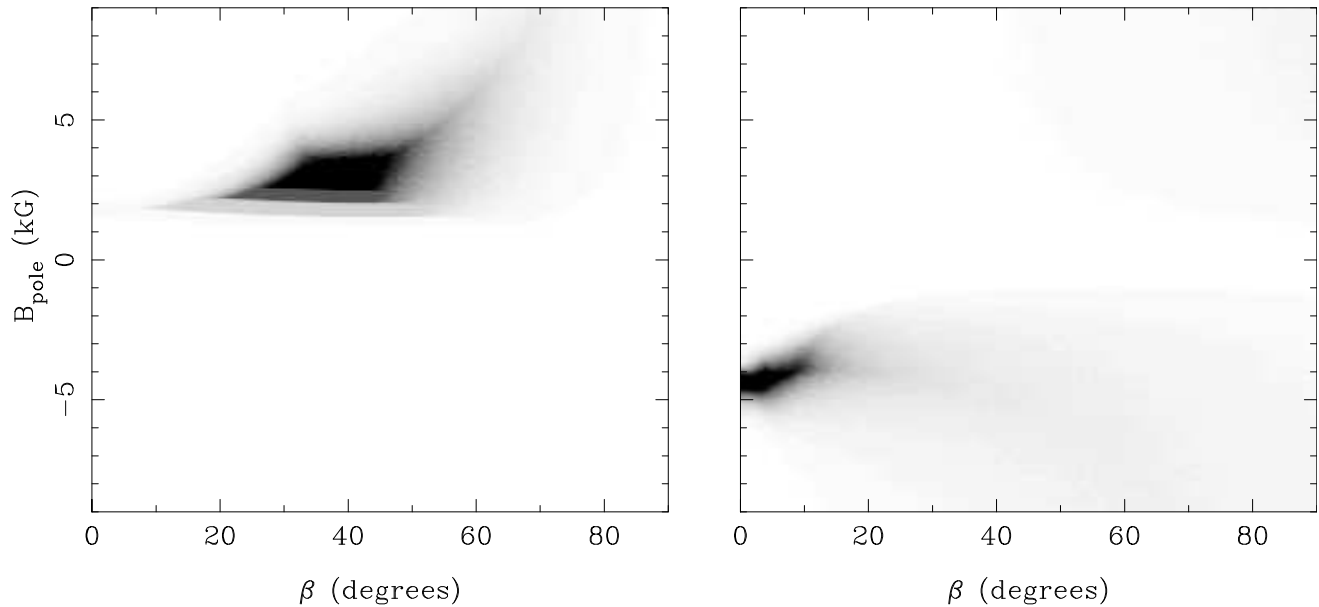}
\end{tabular}

\caption{Kolmogorov-Smirnov statistics for BP~Tau (left) and DF~Tau
(right).  Each pixel shows the probability that the observations are
consistent with a simulated distribution of measurements from a dipole
magnetic field with given values of $B_{\mathrm{pole}}$ and dipole
offset ($\beta$). Black represents a probability of $>50$ per cent,
white is $0$, and the greyscale shows a linear range between these two
values. The step changes seen in the BP~Tau simulation are a
consequence of employing the K-S test with a small number of data
points. } \protect\label{fig:ksprob}

\end{figure*}

Many such sets of observations were simulated for varying values of
$B_{\mathrm{pole}}$ and $\beta$ and a Kolmogorov-Smirnov (K-S) test was carried
out to determine whether the genuine observations were consistent with the
simulated distribution.  In Fig.\ref{fig:ksprob}, the K-S probability is shown as $B_{\mathrm{pole}}$ and $\beta$ are changed, and it can be seen that the models are ruled out for many combinations of the  
these parameters.  In
each case, there is a practical constraint that $B_{\mathrm{pole}}$ must be at
least as great in magnitude as the maximum observed value of $B_z$. Imposing a
narrow range of viewing inclinations will also introduce a degeneracy between
$B_{\mathrm{pole}}$ and $\beta$, i.e. a $B_{\mathrm{pole}}$ value that does
not match the star's actual $B_{\mathrm{pole}}$ might be compensated for by
changing the dipole offset so that the distribution of simulated observations
is more widely spread and then is consistent with the real observations.

The BP~Tau observations are found to be best matched by
$B_{\mathrm{pole}} \sim 3$\,kG and $\beta \sim 40\degr$. DF~Tau was
consistent with $\sim -4.5$\,kG and $\beta < \sim 10\degr$ --
the difference in dipole offset is consistent
with our expectation, since our model gives smaller dipole offsets for
objects with smaller variability for obvious physical reasons.

\section{Conclusions}

Measurements were made of the mean longitudinal magnetic field, $B_z$,
from Zeeman splitting of the He\,{\sc i}~$\lambda 5876$ line in the
spectra of seven classical T~Tauri stars. 
For the star that had been
previously best-studied with this technique, BP~Tau,
the field strengths were mostly within the range found by previous authors \citep{1999ApJ...510L..41J,2003csss...12..729V}, but we also measured the strongest field so far detected for any T~Tauri star (4\,kG).
Another star, DF~Tau,
was found to have a magnetic field that was also within the range of
previous measurements.  A first $B_z$ detection was made for
DN~Tau, and the stars GM~Aur and RW~Aur each showed one
measurement that was above our instrumental sensitivity limit. In total,
five of the seven targets showed evidence for a magnetic field at one or more
epochs.

There are now a total of nine CTTSs in the literature with precise
magnetospheric field measurements, and of this sample a statistically
significant field has been detected in seven objects. Interestingly
our repeat observations of BP~Tau and DF~Tau indicate that the fields
may have characteristics that are persistent on a time-scale of
years (i.e. hundreds of rotation cycles).
It is evident that
classical T~Tauri stars have fields with the magnitude ($\sim1$\,kG) and stability expected  in
theoretical studies of the magnetospheric accretion process (e.g. \citealt{2003ApJ...595.1009R}).

We note that the dipole offset estimates derived for BP~Tau and DF~Tau
(40\degr and $<10$\degr respectively) would result in very different
magnetospheric stream structures according to the
magneto-hydrodynamical models of \cite{2003ApJ...595.1009R}. The
simulations suggest that DF~Tau should accrete from two
diametrically opposed streams, while BP~Tau (with its larger dipole
offset) would have a more complex accretion structure. However, it
appears BP~Tau shows line profile variability that is consistent with
an inclined dipole (\citealt{1996A&A...314..835G}), while the
variability of DF~Tau is apparently difficult to interpret in terms of
the magnetospheric accretion paradigm (\citealt{1997ApJ...474..433J}).

The next stage of these investigations should be a simultaneous
measurement of the mean field and the longitudinal field, perhaps
using time-series measurements to map the surface magnetic field
structure using a technique such as Zeeman Doppler Imaging (ZDI;
\citealt{1989A&A...225..456S}).
\section*{Acknowledgements}

We thank an anonymous referee for their careful reading of the manuscript and
helpful comments.  The WHT is operated on the island of La Palma by the Isaac
Newton Group in the Spanish Observatorio del Roque de los Muchachos of the
Instituto de Astrofisica de Canarias. We thank PATT for the allocation of WHT
time, and the staff of the ING for excellent support.  C.~Johns-Krull is
thanked for providing details of his $B_z$ measurements. RK is funded by PPARC
standard grant PPA/G/S/2001/00081.

\bibliography{nhs} \bibliographystyle{mn2e}

\end{document}